\documentclass[aps,prl,twocolumn,showpacs,superscriptaddress,groupedaddress]{revtex4}

\usepackage{graphicx}  
\usepackage{dcolumn}   
\usepackage{bm}        
\usepackage{amssymb} 

\begin{document}

\newcommand{\be}{\begin{equation}}
\newcommand{\ee}{\end{equation}}
\newcommand{\bn}{\begin{eqnarray}}
\newcommand{\en}{\end{eqnarray}}
\newcommand{\bw}{\begin{widetext}}
\newcommand{\ew}{\end{widetext}}

\draft


\title{Real-space cluster dynamical mean-field approach to the Falicov-Kimball model: An alloy-analogy approach}

\author{P. Haldar, M. S. Laad and S. R. Hassan}\email{{prosenjit|mslaad|shassan}@imsc.res.in}

\address{Institute of Mathematical Sciences, Taramani, Chennai 600113, India \\ and \\
Homi Bhabha National Institute Training School Complex,
Anushakti Nagar, Mumbai 400085, India}

\date{\today}

\widetext

\begin{abstract}
  It is long known that the best single-site coherent potential approximation (CPA) falls short of describing Anderson localization (AL).  Here, we study a binary alloy disorder (or equivalently, a spinless Falicov-Kimball (FK)) model and construct a dominantly analytic cluster extension that treats intra-cluster ($1/d$, $d$=spatial dimension) correlations {\it exactly}.  We find that, in general, the irreducible two-particle vertex exhibits clear non-analyticities {\it before} the band-splitting transition of the Hubbard type occurs, signaling onset of an unusual type of localization at strong coupling.  Using time-dependent response to a sudden local quench as a diagnostic, we find that the long-time wave function overlap changes from a power-law to an anomalous form at strong coupling, lending additional support to this idea.  Our results also imply such novel ``strong'' localization in the equivalent FK model, the simplest {\it interacting} fermion system. 
\end{abstract}


\maketitle

\narrowtext


  Anderson's seminal paper~\cite{pwa-1958} spawned the fertile field of localization in disordered systems.  
While all states in spatial dimension $d=1,2$ are long known to be localized for any arbitrary disorder in 
the ``weak'' localization sense, strong enough disorder is generally expected to lead to exponential localization in all $d$.  In a distinct vein, the {\it exact} and otherwise successful mean-field theory of AL, the coherent potential approximation (CPA) cannot, by construction, describe AL, since it cannot account for coherent backscattering processes that underpin AL.  Nevertheless, CPA has been used in the Vollhardt-W\"olfle (VW) theory to obtain a
phase diagram with AL and metallic phases~\cite{kroha-2015}.   Other schemes marry the CPA with typical medium theory (TMT) to study AL~\cite{dobrosavljevic}.  However, given the necessity of including non-local correlations, 
several heavily numeric-based cluster approaches~\cite{mills,jarrell,rowlands} have also been devised 
with mixed success.  In addition, the simplest model of correlated fermions on a lattice, the Falicov-Kimball model (FKM), is isomorphic to the binary-alloy Anderson disorder model, and exhibits a continuous metal-insulator transition of the Hubbard band splitting type~\cite{hubbardIII}.  One might thus expect the above issues to be relevant for the FKM as well.  To our knowledge, a dominantly {\it analytic} approach to cluster-based techniques in such contexts remains to be attempted, and is potentially of great interest.

   Recent work on many-body localization~\cite{khemani} suggest that at strong disorder the localization length is of the order of lattice constant ($\xi\simeq 1$).  In this limit, an {\it exact} treatment of 
inter-site ``disorder'' ($1/d$) correlations beyond DMFT may thus be adequate to describe ``strong'' localization.  Non-local response to a sudden local quench (a suddenly switched-on localized hole) in this regime exhibits a statistical orthogonality catastrophe (sOC), also studied earlier in the context of correlated impurity potentials in a fermi gas~\cite{altshuler}.
Thus, qualitative change in the long-time response of a system to a sudden local quench, wherein the explicit long-time wave function overlap undergoes a qualitative change at strong disorder, can be a novel diagnostic of strong localization.  Can we study how the long-time response to a sudden local quench evolves across the MIT in the FKM, and can such an endeavor provide deeper insight into ``strong'' localization
at a continuous metal-insulator transition?

  In this paper, we develop an analytic cluster-DMFT for the non-interacting Anderson disorder or FK model

\be
\label{eq:1}
H_{AM}=-t\sum_{\langle i,j\rangle,\sigma}(c_{i\sigma}^{\dag}c_{j\sigma}+h.c) + \sum_{i,\sigma}v_{i}c_{i\sigma}^{\dag}c_{i\sigma}
\ee
on a Bethe lattice.  The $v_{i}$ are random variables with a binary alloy distribution: $P(v_{i})=(1-x)\delta(v_{i}) + x\delta(v_{i}-U)$.  Relabeling $v_{i}=U n_{id}$ where $n_{id}=d_{i}^{\dag}d_{i}$ is the occupation number of a spinless non-dispersive fermion state ($n_{id}=0,1$ for all $i$).  The mapping between the two models implies that the $n_{i,d}$ are randomly distributed according to the above binary alloy distribution.
We extend earlier two-site cluster-DMFT~\cite{laad} (we use the same notations here as in Ref[10]), wherein a crucial advance is 
to go beyond the DMFT-like {\it site}-local Weiss field to one that explicitly incorporates full intra-cluster disorder correlations (see below for details) via a matrix Weiss field.  This is a non-trivial step, and is necessary to obtain the correct causal {\it cluster} propagators 
and self-energies.  A unique and very attractive aspect of our 
c-DMFT is that we obtain explicit closed-form expressions for the cluster propagators (thus self-energies and irreducible charge vertices): 
while $H_{AM}$ is known to be analytically solvable in $d=\infty$~\cite{freericks}, a dominantly analytic cluster extension has remained elusive, 
though the problem has been tackled numerically~\cite{mills,jarrell,rowlands}.  Remarkably, one just needs to solve two coupled non-linear algebraic 
equations for a two-site cluster ($N$ equations for a $N$-site cluster) leading to extreme computational simplification, even with finite alloy 
short-range order (SRO).  This makes it very attractive for use for real correlated systems in conjunction with multiband DMFT or cDMFT.  We will be specifically concerned with studying quantum critical aspects at the the Hubbard type of a continuous MIT accompanied by band splitting.  Extensions to study Anderson localization within the formalism developed in this paper are very interesting, but is deferred for the future.  

\section{MODEL AND SOLUTION}

The Hamiltonian for non-interacting Anderson disorder model or equivalently Falicov Kimball model(FKM) within alloy analogy approximation is:
\begin{equation}
\label{eq:2}
H = -t \sum_{\langle i,j \rangle, \sigma} (c^{\dagger}_{i\sigma}c_{j\sigma} + h.c.)  + \sum_{i\sigma} v_i n_{i\sigma}
\end{equation}  
Here, $v_i$ is taken as diagonal disorder with binary distribution i.e.,
\begin{equation}
\label{eq:3}
P(v_i) = (1-x) \delta(v_i-v_A) + x \delta(v_i-v_B)
\end{equation} 
with $v_A=0$ and $v_B=U$.
We further consider SRO between two nearest neighbour sites   (i,j) as, $f_{ij} = \langle v_i v_j \rangle - \langle v_i  \rangle\langle v_j \rangle= C $, a constant parameter. Although in real materials $f_{ij}$ depends on the $x$, temperature and other physical variables and this dependence should be considered explicitly.

We mapped the Hamiltonian using C-DMFT technique to an effective Anderson impurity model with impurity as
a two site cluster embedding by an effective dynamical bath.

The Hamiltonian for the Anderson Impurity Model is given as:
\bn
\label{eq:4}
H_{imp} = -t \sum_{\sigma}(c^{\dagger}_{0\sigma}c_{\alpha\sigma} +h.c.) +   U\sum_{i\in \{ 0,\alpha \} \sigma} x_i n_{i\sigma} \nonumber\\
 + \sum_{i\in \{ 0,\alpha \},k,\sigma} (v_{ki}c^{\dagger}_{i\sigma}c_{k\sigma} + h.c.) + \sum_{k,\sigma}  \epsilon_{k}  c^{\dagger}_{k\sigma} c_{k\sigma} 
\en
The first term is the hopping between two sites ($0,\alpha$) of the cluster impurity, second term corresponds to the interaction of the impurity, third term describes the hybridization between the impurity and the bath and fourth term describes the dispersive bath. Here, $x_i \equiv n_{id}$ is the occupation of localized fermions in FKM.

Two-site cluster impurity Green's function is given in matrix form:
\[ \hat{\mathbf{G}}= \left(  \begin{array}{cc}
G_{00}(\omega) & G_{\alpha 0}(\omega) \\
G_{\alpha 0}(\omega) & G_{00}(\omega) \end{array} \right) \] 
Here, the element of $\hat{\mathbf{G}}$ is defined as, $G^{\sigma}_{ij}(\omega):= \langle c_{i\sigma};c^{\dagger}_{j\sigma}\rangle$  with $\sigma$ is the spin indices, i.e. $\sigma\in \{ \uparrow, \downarrow\}$. 

The Equation of Motion (EOM) for $G^{\sigma}_{ij}(\omega)$ is:
\bn
\label{eq:6}
\omega G_{ij}^{\sigma}(\omega)=
\delta_{ij}-t \sum_{l\neq i}G_{lj}^{\sigma}(\omega)+
U<x_{i}c_{i\sigma};c_{j\sigma}^{\dag}>\nonumber\\
 + \sum_k v_{ki} G_{kj}^{\sigma}(\omega)
\en
Where we use the identity for fermions,
$\omega\langle \hat{A};\hat{B}\rangle = \langle[\hat{A},\hat{B}]_{+}\rangle + \langle [[\hat{A},\hat{H}_{imp}];\hat{B}]_{+} \rangle $. 
Similarly, the EOM for higher order Green's function $\langle x_0 c_{i\sigma};c_{j\sigma}^{\dag}\rangle$ is:
\bn
\label{eq:7}
\omega \langle x_0 c_{i\sigma};c_{j\sigma}^{\dag}\rangle = \langle x_0 \rangle \delta_{ij} - t \sum_{l\neq i}\langle x_0 c_{l\sigma};c^{\dag}_{j\sigma}\rangle + U \langle x_0 x_i c_{i\sigma};c^{\dag}_{j\sigma}\rangle \nonumber \\
+ \sum_k v_{ki} \langle  x_0 c_{k\sigma};c^{\dag}_{j\sigma}\rangle
\en
and EOM for $\langle x_{\alpha} c_{i\sigma};c_{j\sigma}^{\dag}\rangle$ is:
\bn
\label{eq:8}
\omega \langle x_{\alpha} c_{i\sigma};c_{j\sigma}^{\dag}\rangle = \langle x_{\alpha} \rangle \delta_{ij} - t \sum_{l\neq i}\langle x_{\alpha} c_{l\sigma};c^{\dag}_{j\sigma}\rangle + U \langle x_{\alpha} x_i c_{i\sigma};c^{\dag}_{j\sigma}\rangle \nonumber \\
+ \sum_k v_{ki} \langle  x_{\alpha} c_{k\sigma};c^{\dag}_{j\sigma}\rangle
\en

Again, EOM for $<x_0 x_{\alpha}c_{i\sigma};c_{j\sigma}^{\dag}>$ is:
\bn
\label{eq:9}
(\omega-U)<x_{0}x_{\alpha}c_{i\sigma};c_{j\sigma}^{\dag}> = \langle x_0 x_{\alpha}\rangle \delta_{ij}
-t\sum_{l\neq i} <x_{0}x_{\alpha}c_{l\sigma};c_{j\sigma}^{\dag}> \nonumber\\
+ \sum_{k} v_{ki}<x_{0}x_{\alpha}c_{k\sigma};c_{j\sigma}^{\dag}>\nonumber\\
\en
Here, $\langle x_0 x_{\alpha}\rangle=\langle x_{\alpha} x_0\rangle\equiv \langle x_{0\alpha}\rangle$.\\
EOM for the $\langle A_{0\alpha}c_{k\sigma};c^{\dag}_{j\sigma}\rangle$ is:
\be
\label{eq:10}
(\omega - \epsilon_k)\langle A_{0\alpha}c_{k\sigma};c^{\dag}_{j\sigma}\rangle = \sum_i v^{*}_{ki} \langle A_{0\alpha}c_{i\sigma};c^{\dag}_{j\sigma}\rangle
\ee
Here, $A_{0\alpha}\equiv 1,x_0,x_{\alpha},x_{0\alpha}$.\\
We derive the generalized form of cluster Green's function by solving (\ref{eq:6}-\ref{eq:10}),
\bw
\begin{eqnarray}
\label{eq:11}
G_{ij}(\omega) = \left[\frac{1-\langle x_0\rangle-\langle x_{\alpha}\rangle+\langle x_{0\alpha}\rangle}{\xi_2(\omega)}+\frac{\langle x_0\rangle-\langle x_{0\alpha}\rangle}{\xi_2(\omega)-U} \right] \left[  \delta_{ij} - \frac{F_2(\omega)}{(t-\Delta_{\alpha 0}(\omega))}(1-\delta_{ij}) \right] \nonumber\\
+\left[\frac{\langle x_{\alpha}\rangle-\langle x_{0\alpha}\rangle}{\xi_1(\omega)}+\frac{\langle x_{0\alpha}\rangle}{\xi_1(\omega)-U} \right] \left[  \delta_{ij} - \frac{F_1(\omega)}{(t-\Delta_{\alpha 0}(\omega))}(1-\delta_{ij}) \right]
\end{eqnarray}
\ew
Here, $\xi_1(\omega)=(\omega-\Delta_{00}(\omega)-F_1(\omega))$, $\xi_2(\omega)=(\omega-\Delta_{00}(\omega)-F_2(\omega))$ and $F_{1}(\omega) \equiv \frac{(t-\Delta_{\alpha 0}(\omega))^2}{\omega-\Delta_{00}(\omega)-U}$,  
$F_{2}(\omega)\equiv \frac{(t-\Delta_{\alpha 0}(\omega))^2}{\omega-\Delta_{00}(\omega)}$. 
We obtain renormalized $\xi_{1(2)}$ and  $t$  in the diagonal Green's function ( $G_{00}(\omega)$ ) as compared with the results of \cite{laad}. $\xi_{1(2)}$ and $t$ are renormalized by  $\tilde{\xi}_{1(2)}=(\xi_{1(2)}-\Delta_{00}(\omega))$ and $\tilde{t}=(t-\Delta_{\alpha 0}(\omega))$ respectively.   
The bath function $\hat{\mathbf{\Delta}}(\omega)$ in the two sites cluster model is a 2x2 matrix,
\[ \hat{\mathbf{\Delta}}(\omega) = \left( \begin{array}{ccc}
 \Delta_{00}(\omega) & \Delta_{\alpha 0}(\omega)  \\
\Delta_{\alpha 0}(\omega) & \Delta_{00}(\omega)  \end{array} \right),\] Here, $\hat{\mathbf{\Delta}}(\omega)$ is computed from matrix generalization of the dynamic Weiss 
field~\cite{freericks}, 

\be
\label{eq:12}
G(\omega)=\int_{-W}^{+W}\frac{\rho_{0}(\epsilon) 
d\epsilon}{G^{-1}(\omega)+\Delta(\omega)-\epsilon} \;.
\label{eq18}
\ee
where $\rho_{0}(\epsilon)$ is the unperturbed DOS.


\section{General Formalism for the Two-site Cluster Method}

   We can also exactly estimate the cluster irreducible vertex functions and (charge) susceptibility by generalizing the well-known 
procedure employed in DMFT studies~\cite{freericks}.  It turns out that exploiting cluster symmetry is particularly useful in this instance, and markedly simplifies the analysis.
The solution of the two site cluster impurity problem gives the following matrix Green function and the self energy,
\[ \hat{\mathbf{G}} = \left( \begin{array}{ccc}
 G_{00} & G_{\alpha 0}  \\
G_{\alpha 0} & G_{00}  \end{array} \right), 
\hat{\mathbf{\Sigma}} = \left( \begin{array}{ccc}
 \Sigma_{00} & \Sigma_{\alpha 0}  \\
\Sigma_{\alpha 0} & \Sigma_{00}  \end{array} \right),\]
To go over to a representation where these matrices are diagonal in the cluster momenta, $\mathbf{K}$ points, $K_{I}=(0,0,..)$ and $K_{II}=(\pi,\pi,..)$. we divide the Brillouin zone into two sub-zones as done in CDMFT studies for the Hubbard model~\cite{liebsch}.  For brevity, we label Region I and II by S and P respectively.  Now, self energy and Green function matrices take on the diagonalized forms,
\[ \hat{\mathbf{G}} = \left( \begin{array}{ccc}
 G_{S} & 0  \\
0 & G_{P}  \end{array} \right), 
\hat{\mathbf{\Sigma}} = \left( \begin{array}{ccc}
 \Sigma_{S} & 0  \\
0 & \Sigma_{P}  \end{array} \right),\]
where,
\bn
\nonumber
G_{S}=G_{00}+G_{\alpha 0},\hspace{0.4 cm}
G_{P}=G_{00}-G_{\alpha 0}
\en
and,
\bn
\nonumber
\Sigma_{S}=\Sigma_{00}+\Sigma_{\alpha 0},\hspace{0.4 cm}
\Sigma_{P}=\Sigma_{00}-\Sigma_{\alpha 0}
\en
with,
\be
G_{S(P)}(\omega) = \int \frac{\rho^0_{S(P)}(\epsilon)d\epsilon}{\omega+\mu-\epsilon -\Sigma_{S(P)}}
\ee
The partial density of states, which are now nothing else than the ${\bf K}$-dependent spectral functions, are given by,
\be
\rho^0_{S(P)}(\epsilon) = 2\times \int_{\mathbf{k}\in S(P)} d\mathbf{k} \delta(\epsilon-\epsilon_k)
\ee
  In this representation, it turns out to be easier to compute the irreducible vertex functions.  Specifically, the {\it only} quantity relevant for the disorder problem is the irreducible particle-hole (p-h) vertex function $\hat{\mathbf{\Gamma}}$ is given as $\hat{\mathbf{\Gamma}}=\frac{\delta\hat{\mathbf{\Sigma}}}{\delta\hat{\mathbf{G}}}$ (the ``spin fluctuation'' and ``pairing'' vertex appear in the FKM, but are irrelevant for the disorder problem).
Since both self energy and Green function matrices are diagonal in the S(P) basis, the vertex function is separable with respect to the S or P channel as well:
\be
\Gamma_{S(P)}=\frac{\delta\Sigma_{S(P)}}{\delta G_{S(P)}}
\ee
\section{Calculation of Charge Susceptibility}

   With explicit knowledge of the p-h irreducible vertex as above, the momentum-dependent susceptibility corresponding to the S(P) channels is evaluated using the Bethe-Salpeter equation (BSE),
\bw
\bn
\nonumber
 \chi_{S(P)}(\textbf{q},i\omega_m,i\omega_n;i\nu_l) = \chi^{0}_{S(P)}(\textbf{q},i\omega_m;i\nu_l)\delta_{mn} - T\sum_{n'} \chi^{0}_{S(P)} (\textbf{q},i\omega_m;i\nu_l)
 &\times \Gamma_{S(P)}(i\omega_m,i\omega_{n'};i\nu_l)\chi_{S(P)}(\textbf{q},i\omega_{n'},i\omega_n;i\nu_l)\\
\label{eq23}
\en
\ew
To make progress, we proceed along lines similar to those adopted in DMFT studies~\cite{freericks}.
\vspace{0.3cm}

$(i)$ The full susceptibility is found by summing over the fermionic Mastubara frequencies, $\chi_{S(P)}(\textbf{q},i\nu_l)=T\sum_{mn}\chi_{S(P)}(\textbf{q},i\omega_m,i\omega_n;i\nu_l)  $. 
\vspace{0.3cm}
 
$(ii)$ the vertex functions $\Gamma_{S(P)}(i\omega_m,i\omega_{n'};i\nu_l)$ are evaluated as,
\be
 \Gamma_{S(P)}(i\omega_n,i\omega_{m};i\nu_l) = \frac{1}{T} \frac{\Sigma^{S(P)}_n-\Sigma^{S(P)}_{n+l}}{G^{S(P)}_n - G^{S(P)}_{n+l}} \delta_{m,n}
\label{eq24}
\ee
As, $\chi_{S(P)}$ are diagonal in S or P channel, we keep only the channel index $S$, with the understanding that an identical calculation holds for the $P$ channel.
Using $\Gamma $ from equation(27) we find,
\bn
\nonumber
 \chi^{S}(\textbf{q},i\omega_m,i\omega_n;i\nu_l)& = &\chi_{0}^{S}(\textbf{q},i\omega_m;i\nu_l)\delta_{mn} - T  \chi_{0}^{S} (\textbf{q},i\omega_m;i\nu_l)\\
\nonumber
 &\times& \Gamma^{S}(i\omega_m,i\omega_{m};i\nu_l)\chi^{S}(\textbf{q},i\omega_{m},i\omega_n;i\nu_l)\\
\label{equ25}
\en

\be
\Rightarrow \chi^{S}(\textbf{q},i\omega_m,i\omega_n;i\nu_l)=  \frac{\chi_{0}^{S}(\textbf{q},i\omega_m;i\nu_l)\delta_{mn}}{1+\chi_{0}^{S}(\textbf{q},i\omega_m;i\nu_l)\frac{\Sigma_m^{S}-\Sigma_{m+l}^{S}}{G_m^{S}-G_{m+l}^{S}}}
\label{equ26}
\ee

\vspace{0.3cm}

$(iii)$ Now, the full lattice susceptibility with $\textbf{q}$ replaced by $X(\textbf{q})$ is given by,
\bn
\nonumber
 \chi^{S}(X,i\nu_l\neq0)&=&T\sum_{m,n} \chi^{S}(X,i\omega;i\nu_l) \\
\nonumber
&=& T\sum_{m,n}  \frac{\chi_{0}^{S}(X,i\omega_m;i\nu_l)\delta_{mn}}{1+\chi_{0}^{S}(X,i\omega_m;i\nu_l)\frac{\Sigma_m^{S}-\Sigma_{m+l}^{S}}{G_m^{S}-G_{m+l}^{S}}} \\
\label{equ27}
\en
where, $X(\mathbf{q})=\lim_{d\to \infty}\sum_{i=1}^{d} cos(\frac{q_i}{d})$
\be
\Rightarrow \chi^{S}(X,i\nu_l\neq0)=T\sum_{m}  \frac{\chi_{0}^{S}(X,i\omega_m;i\nu_l)}{1+\chi_{0}^{S}(X,i\omega_m;i\nu_l)\frac{\Sigma_m^{S}-\Sigma_{m+l}^{S}}{G_m^{S}-G_{m+l}^{S}}}
\label{equ28}
\ee 
\\
$(iv)$ For $\textbf{q}=0(X=1)$, we find that,
\be
 \chi(1;i\nu_l\neq0) = -T \sum_{m} \frac{G_m-G_{m+l}}{i\nu_l} =0
\label{equ29}
\ee

This just reflects conservation of the total $c$-fermion number, and thus vanishes by symmetry.
\vspace{0.3cm}

$(v)$ For generic $\textbf{q} (X=0)$, we calculate the sum over mastubara frequency by contour integration. The bare susceptibility for 
X=0 is given as,$\chi_0^{S}(0,i\omega_m;\nu_l)=-G_m^{S}G_{m+l}^{S}$.
\be
\Rightarrow \chi^{S}(X=0,i\nu_l\neq0)=-T\sum_{m}  \frac{G_m^{S}G_{m+l}^{S}}{1-G_m^{S}G_{m+l}^{S}\frac{\Sigma_m^{S}-\Sigma_{m+l}^{S}}{G_m^{S}-G_{m+l}^{S}}}
\label{equ28}
\ee 
As mentioned before, a similar analysis holds for the $P$-channel at every step in the procedure above.

After performing analytical continuation from mastubara frequency to real frequency in the standard way, the final
expression for the susceptibility corresponding to the S or P channel becomes,
\bw
\bn
\nonumber
\chi_{S(P)}(X=0,\nu \neq 0)=& \frac{1}{2 \pi i} \int_{-\infty}^{\infty} d\omega \{ f(\omega) \frac{G_{S(P)}(\omega)G_{S(P)}(\omega+\nu)}{1-G_{S(P)}(\omega)G_{S(P)}(\omega+\nu)[\Sigma_{S(P)}(\omega)-\Sigma_{S(P)}(\omega+\nu)]/(G_{S(P)}(\omega)-G_{S(P)}(\omega+\nu))}  \\ 
\nonumber
&- f(\omega+\nu) \frac{G_{S(P)}^{*}(\omega)G_{S(P)}^{*}(\omega+\nu)}{1-G_{S(P)}^{*}(\omega)G_{S(P)}^{*}(\omega+\nu)[\Sigma_{S(P)}^{*}(\omega)-\Sigma_{S(P)}^{*}(\omega+\nu)]/(G_{S(P)}^{*}(\omega)-G_{S(P)}^{*}(\omega+\nu))}  \\ 
\nonumber
&-[f(\omega)- f(\omega+\nu)] \frac{G_{S(P)}^{*}(\omega)G_{S(P)}(\omega+\nu)}{1-G_{S(P)}^{*}(\omega)G_{S(P)}(\omega+\nu)[\Sigma_{S(P)}^{*}(\omega)-\Sigma_{S(P)}(\omega+\nu)]/[G_{S(P)}^{*}(\omega)-G_{S(P)}(\omega+\nu)]} \} \\ 
\label{equ31}
\en
\ew
 where we have replaced the retarded green function $G^R_{S(P)}$ by $G_{S(P)}$ and the advanced green function $G^A_{S(P)}$ by the complex conjugate of the retarded Green function $G^*_{S(P)}$.  

  The procedure (i)-(v) allows us to {\it exactly} estimate the cluster p-h irreducible vertices and the charge susceptibility in the FKM to $O(1/d)$.   

\vspace{0.5cm}

\section{Results}

\vspace{1.0cm}

{\bf One-Particle Spectral Response}

\vspace{0.5cm}

   We now present our results.  We work with a semicircular unperturbed density of states (DOS) as appropriate for a Bethe lattice in high-$d$, given by 
$\rho_{0}(\epsilon)=(2/\pi D)\sqrt{D^{2}-\epsilon^{2}}$ where $D=2t=1$ is the $c$-fermion half-band width.  We begin with a ``particle-hole symmetric'' case with 
$\langle n_{ic}\rangle=1/2$ and a probability distribution, $P(v_{i})=1/2[\delta(v_{i})+\delta(v_{i}-U)]$ or, in the FKM context with a half-filled $c$-fermion 
band.
\begin{figure}
\includegraphics[width=1.\columnwidth , height= 
1.\columnwidth]{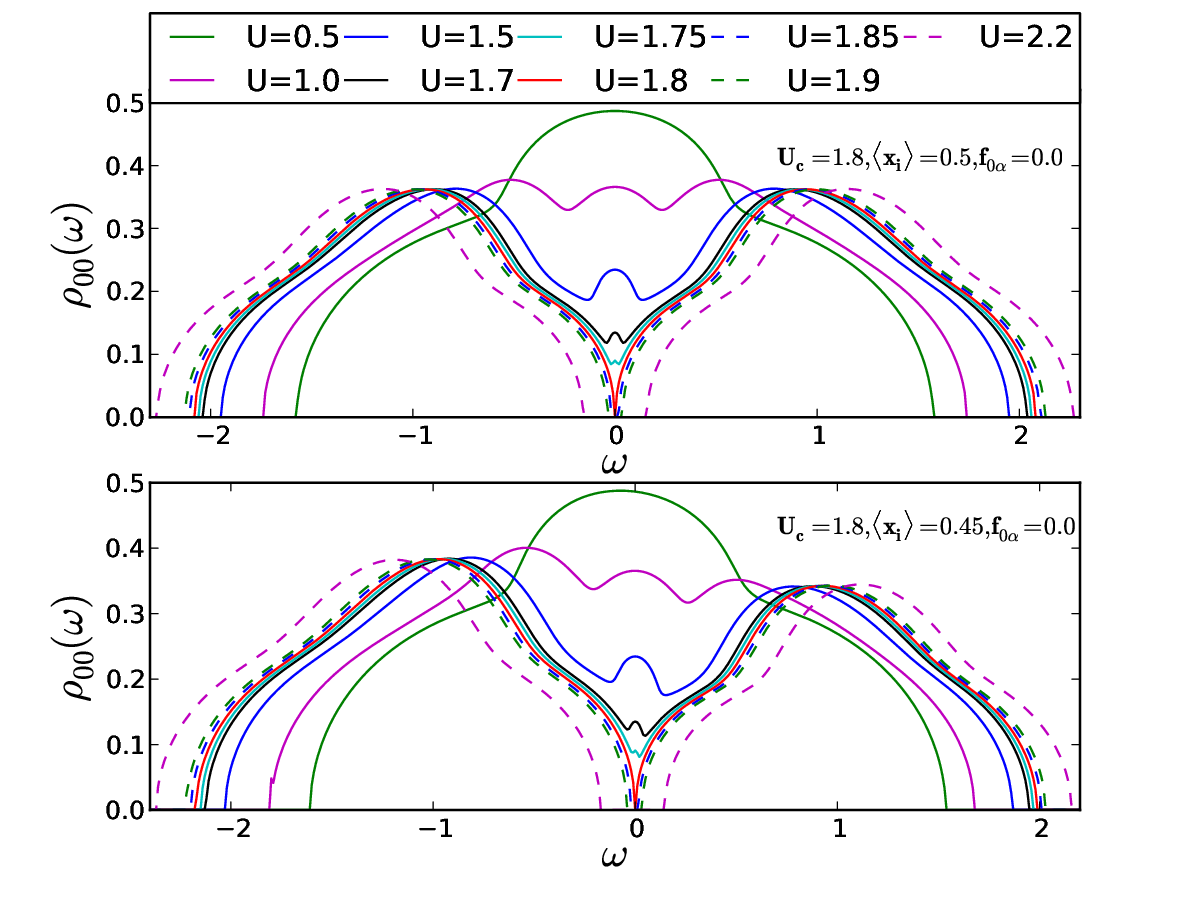} 
\caption{(Color online)  The local density of states (LDOS) of the binary-alloy disorder model for p-h symmetry (upper panel) and p-h asymmetric case (lower panel).  A clear continuous band-splitting transition of the Hubbard (or Falicov-Kimball model-like) variety is seen in both cases.  At $U_{c}=1.8$ (red curve), the LDOS exhibits a critical $|\omega|^{1/3}$ singular behavior in both cases.}
\label{fig:fig1}
\end{figure}  
In Fig.~\ref{fig:fig1}, we show the local DOS (LDOS) as $U$ is increased through a critical $U_{c}=1.8$, where a {\it continuous} metal-insulator 
transition (MIT) of the Mott-Hubbard type occurs via the Hubbard-band splitting.  Comparing with the exact DMFT solution~\cite{freericks}, we see that 
incorporation of dynamical effects of $1/d$ correlations in our two-site CDMFT gives rise to additional features in the LDOS.  These features arise from 
repeated scattering of the electrons off spatially separated scattering centers and are visible even for the totally random case, defined as 
$f_{0\alpha}=\langle x_{0}x_{\alpha}\rangle -\langle x_{0}\rangle\langle x_{\alpha}\rangle =0$.  In the lower panel, we show the LDOS for the asymmetric FKM, 
with $\langle n_{i,d}\rangle =\langle x_{i}\rangle =0.45$, wherein loss of particle-hole symmetry is faithfully reflected as an asymmetric LDOS.  It is clear 
that the MIT is associated with a genuine quantum critical point (QCP). 
   The advantage of CDMFT is that cluster {\it spectral} functions, defined as $A({\bf K},\omega)$ with ${\bf K}=(0,0),(\pi,\pi)$ can be explicitly read off.  
\begin{figure}
\includegraphics[width=1.\columnwidth , height= 
1.\columnwidth]{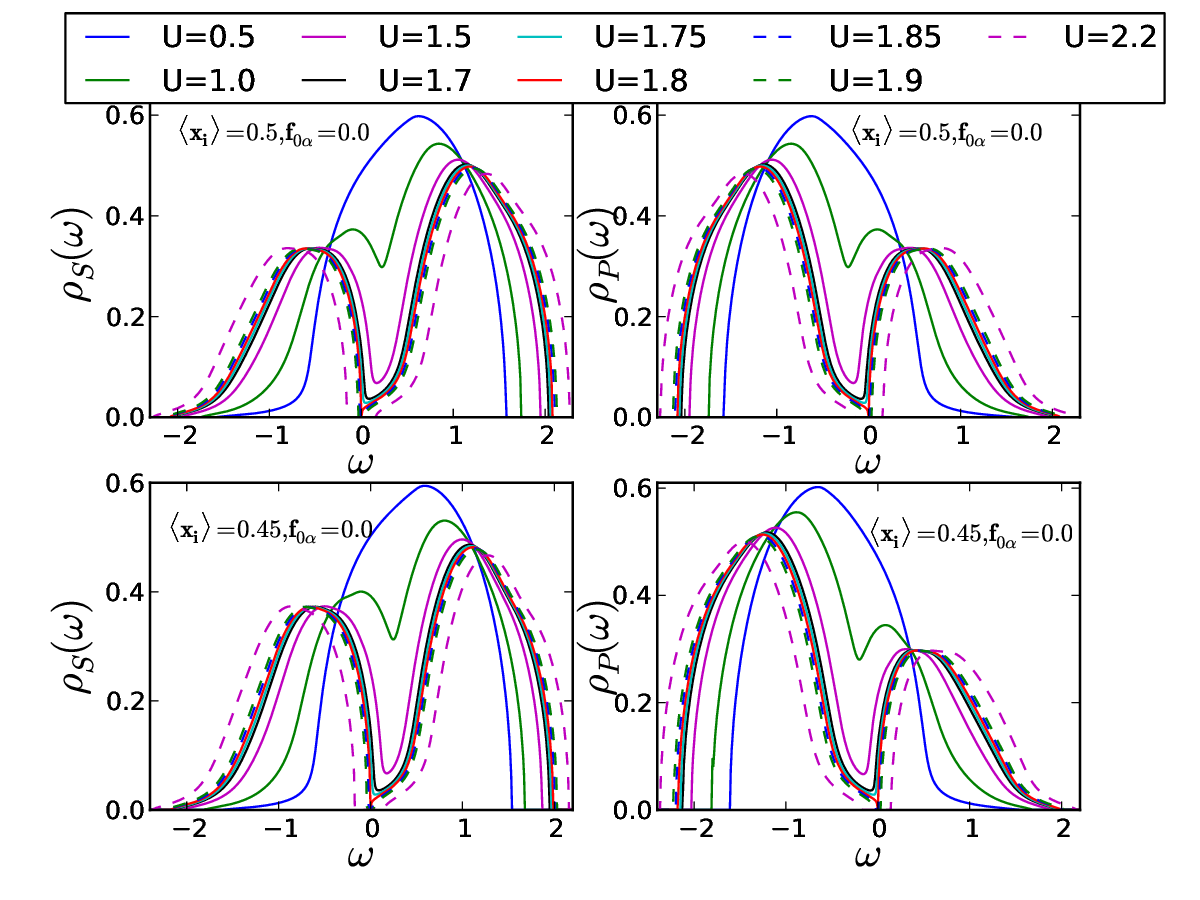} 
\caption{(Color online) The cluster-momentum resolved one-electron spectral functions for same parameters as in Fig.~\ref{fig:fig1}. For the p-h symmetric case, the symmetry relation $\rho_{S}(\omega)=\rho_{P}(-\omega)$ is clearly satisfied as it must be (upper panel).}
\label{fig:fig2}
\end{figure}
In Fig.~\ref{fig:fig2}, we exhibit $A({\bf K},\omega)$ as a function of $U$.  It is obvious that $\rho_{S}(\omega)=A({\bf K}=(0,0),\omega)=\rho_{P}(-\omega)=A({\bf K}=(\pi,\pi),-\omega)$ for the 
particle-hole symmetric case, as it must be.  For the Bethe lattice, we find that the LDOS, $\rho(\omega)=C|\omega|^{1/3}$ (shown in Fig.~\ref{fig:fig11}) exactly at the QCP ($U_{c}=1.8$), 
a result similar to that found for the same model in DMFT~\cite{vandongen}.  The spectral functions also exhibit these singular features, albeit in a 
${\bf K}$-dependent fashion.  Notwithstanding these similarities, we stress that our extension of DMFT faithfully captures the feedback of the non-local 
(intracluster) correlations on the single-particle DOS and the self-energies (see below) in contrast to DMFT, where such $1/d$ feedback effects are absent.
\begin{figure}
\includegraphics[width=1.\columnwidth , height= 
1.\columnwidth]{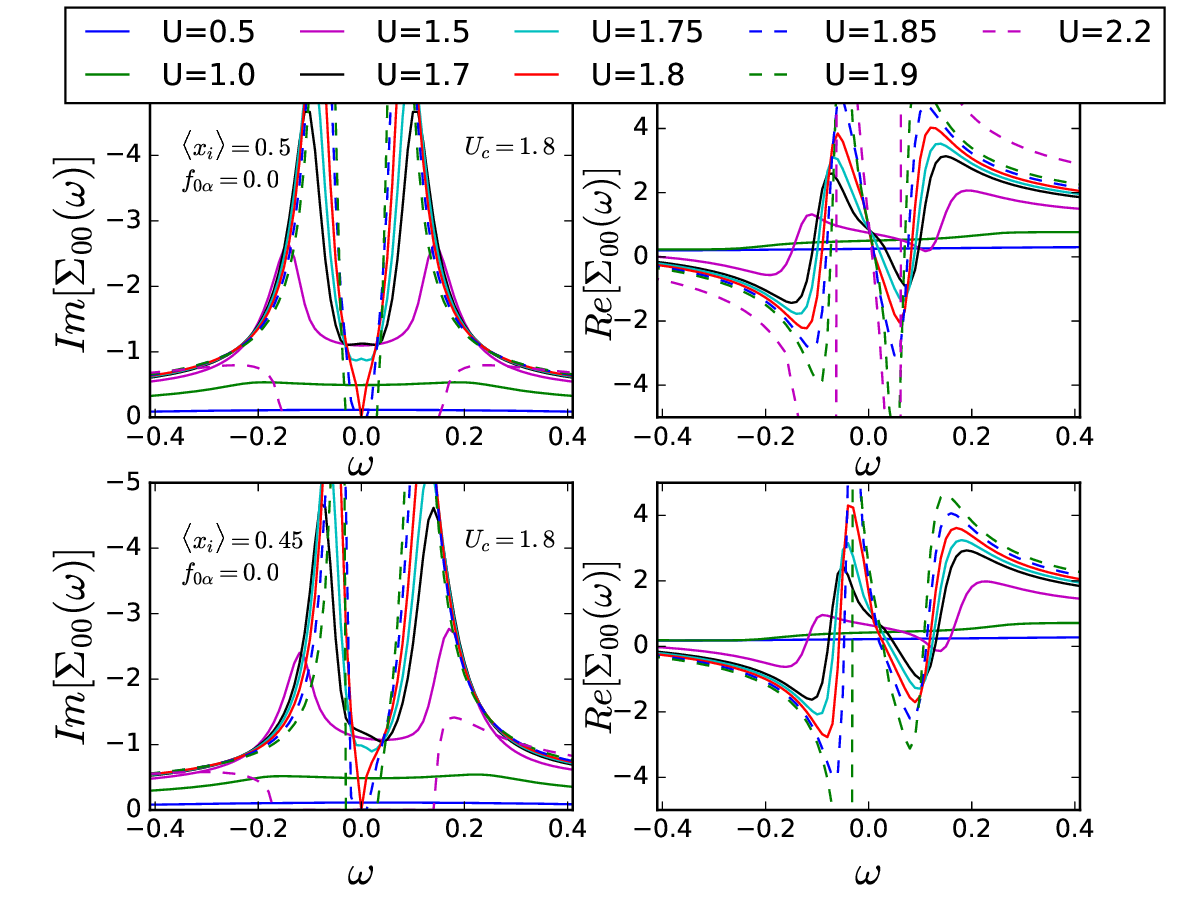} 
\caption{(Color online) $\Sigma_{00}(\omega)$ (both real and imaginary part) vs $U$ for the binary-ally disorder problem for the same parameters as in Fig.~\ref{fig:fig1}.  For small $U$, our results agree with self-consistent Born approximation (constant Im$\Sigma_{00}(\omega)$).  As $U$ increases, Im$\Sigma_{00}(\omega)$ develops marked low-energy structure, and at $U_{c}=1.8$ (red curve), Im$\Sigma_{00}(\omega)\simeq |\omega|^{1/3}$, reflecting the non-perturbative nature of the ``Hubbard III'' quantum criticality. The $Re\Sigma_{00}(\omega)$ shows discontinuity at $\omega=0$ at the critical U.}
\label{fig:fig3}
\end{figure}
   In the left panels of Fig.~\ref{fig:fig3}, we exhibit the imaginary part of the cluster-local self-energy, Im$\Sigma_{00}(\omega)$ for the same parameter values as above. 
 For small $U$, Im$\Sigma_{00}$ weakly depends on $\omega$, and is sizable only near $\omega=0$.  However, it has the {\it wrong} sign, $i.e$, a minimum, instead
of a maximum characteristic of a Landau Fermi liquid, at $\omega=0$.  Thus, the metallic state is incoherent and {\it not} a Landau Fermi liquid (LFL).
This is again a feature in common with DMFT studies.  In DMFT, it is well known that this 
feature becomes more prominent as $U$ increases, and diverges at the MIT~\cite{freericks}.  In CDMFT, however, Im$\Sigma_{00}$ develops marked structure already
at $U=1.0$: it develops a maximum at $\omega=0$, which progressively sharpens up with increasing $U$ in the incoherent metallic regime. 
\begin{figure}
\includegraphics[width=1.\columnwidth , height= 
1.\columnwidth]{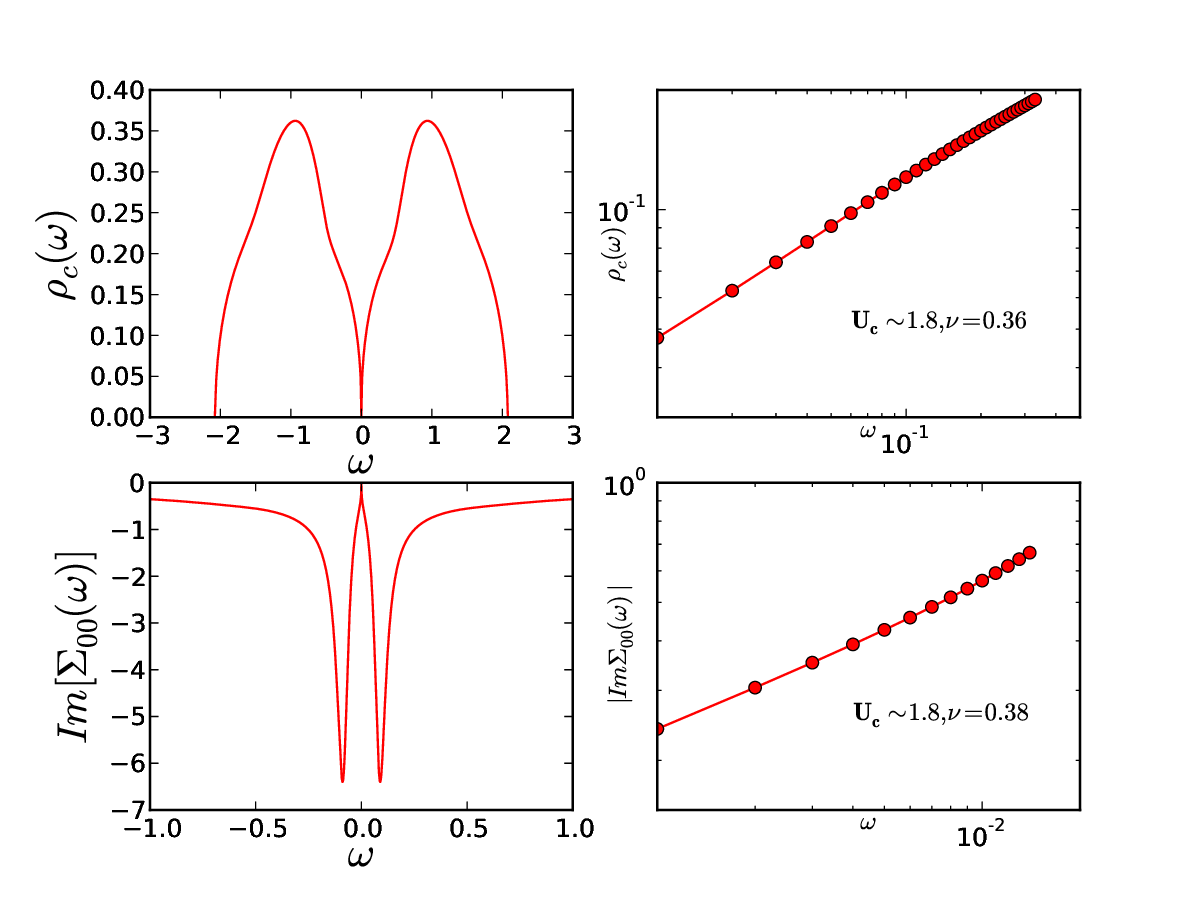} 
\caption{(Color online) Exponent of $\rho_{00}(\omega)$ and $Im\Sigma_{00}(\omega)$ closed to the Fermi energy at critical value of U with symmetric alloy}
\label{fig:fig11}
\end{figure}
Interestingly,
$(i)$ right at $U_{c}$ (red curve), Im$\Sigma_{00}(\omega)=c|\omega|^{1/3}$ (shown in Fig.~\ref{fig:fig11}), reminiscent of what is expected in a {\it power-law liquid}, in strong contrast 
to what happens in DMFT, where it diverges. The real part of local self energy (Re$\Sigma_{00}(\omega)$) is shown in the right panels of Fig.~\ref{fig:fig3}. For p-h symmetric case (shown in upper right panel of Fig.~\ref{fig:fig3}), Re$\Sigma_{00}(\omega)$ is U/2.0  at $\omega$=0 for all values of U. If we see $Re[\Sigma_{00}(\omega)]-\frac{U}{2}$ it changes sign according to the $\omega$ near the Fermi level and at the transition point ($U\sim U_c$) it shows steep discontinuity at $\omega=0$. The source of gap opening comes from the divergence of $\frac{\partial}{\partial \omega} Re\Sigma_{00}(\omega)$ at $\omega$=0. For $ U > U_{c}$, opening up of a ``Mott'' gap in the LDOS goes hand-in-hand with the divergence of $\frac{\partial}{\partial \omega} Re\Sigma_{00}(\omega)$ and vanishing Im$\Sigma_{00}(\omega)$ in the gap.  In all cases, we also find power-law fall-off in self-energies at high energy
and, more interestingly, clear isosbestic points (where Im$\Sigma_{00}(\omega)$ is independent of $\omega$) at $\Omega=\pm 0.2t$.  We also find (see lower panels of Fig.~\ref{fig:fig3}) 
that moving away from p-h symmetry does not 
qualitatively change the above features. 
\begin{figure}
\includegraphics[width=1.\columnwidth , height= 
1.\columnwidth]{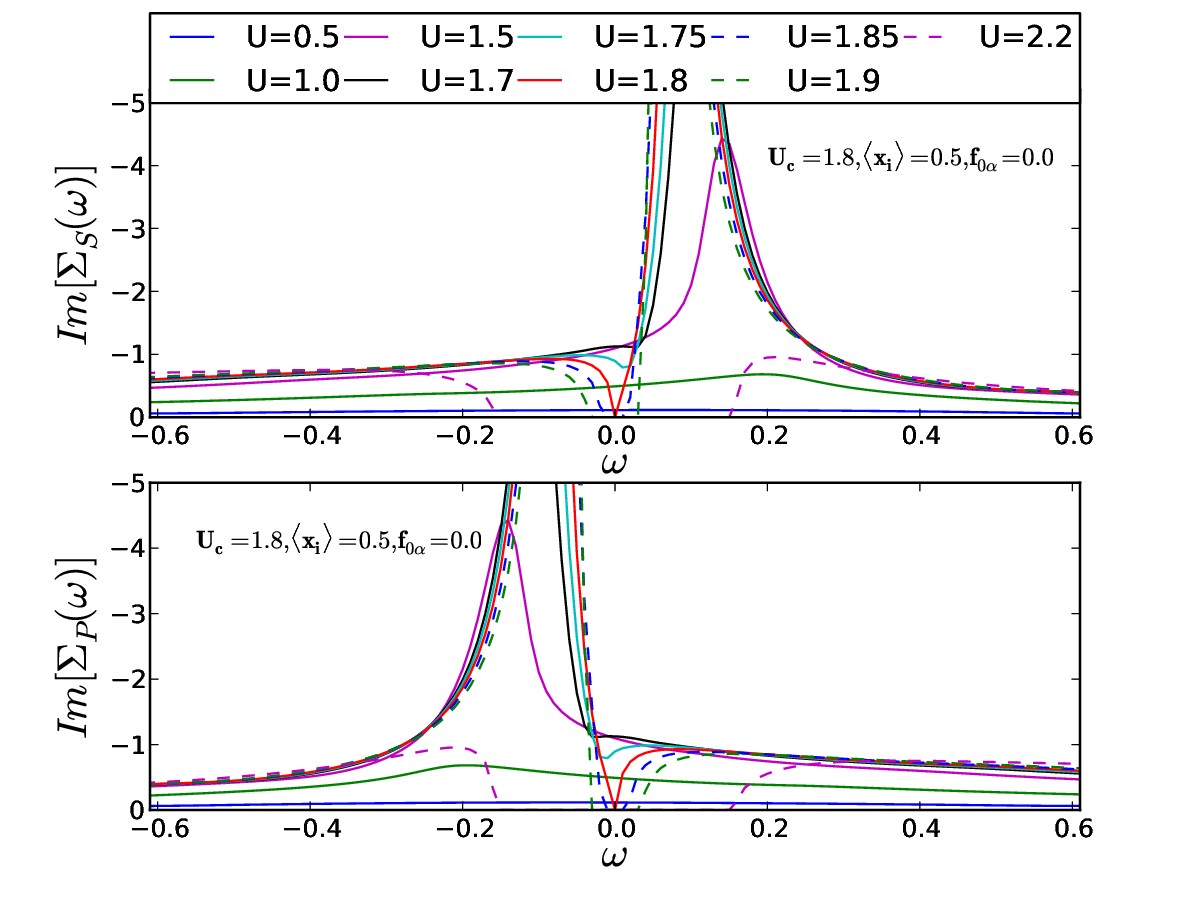} 
\caption{(Color online) Same as Fig.~\ref{fig:fig3}, but now for the cluster-momentum resolved self-energies.  It is clear that the symmetry relation Im$\Sigma_{S}(\omega)=$Im$\Sigma_{P}(-\omega)$ holds in the p-h symmetric case.}
\label{fig:fig4}
\end{figure}
 
\begin{figure}
\includegraphics[width=1.\columnwidth , height= 
1.\columnwidth]{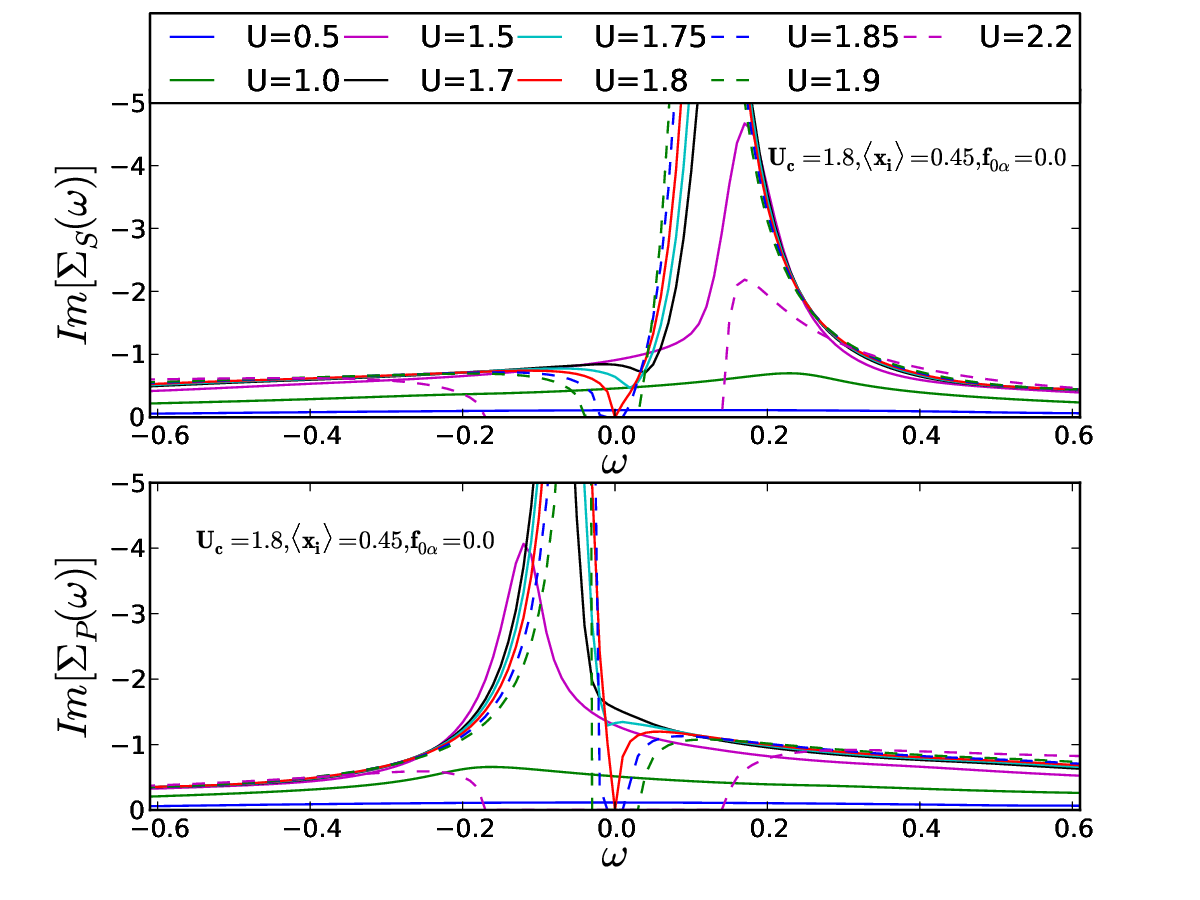} 
\caption{(Color online) Same as in Fig.~\ref{fig:fig4}, but for the p-h asymmetric case.  Though no symmetry is expected nor found here, the critical features are unaffected, since Im$\Sigma_{S,P}(\omega)$ indeed exhibit the
same non-analytic feature ($\simeq |\omega|^{1/3}$ behavior) for $\omega <0 (S)$ and $\omega >0 (P)$.}
\label{fig:fig5}
\end{figure}

Finally, CDMFT allows a direct evaluation of the ${\bf K}$-dependent self-energies, which we exhibit 
in Fig.~\ref{fig:fig4} and Fig.~\ref{fig:fig5}.  
As a cross-check, we find that Im$\Sigma({\bf K}=(0,0),\omega)=$Im$\Sigma({\bf K}=(\pi,\pi),-\omega)$, as 
required by particle-hole (p-h) symmetry for $\langle n_{i,d}\rangle=0.5$.

\begin{figure}
\includegraphics[width=1.\columnwidth , height= 
1.\columnwidth]{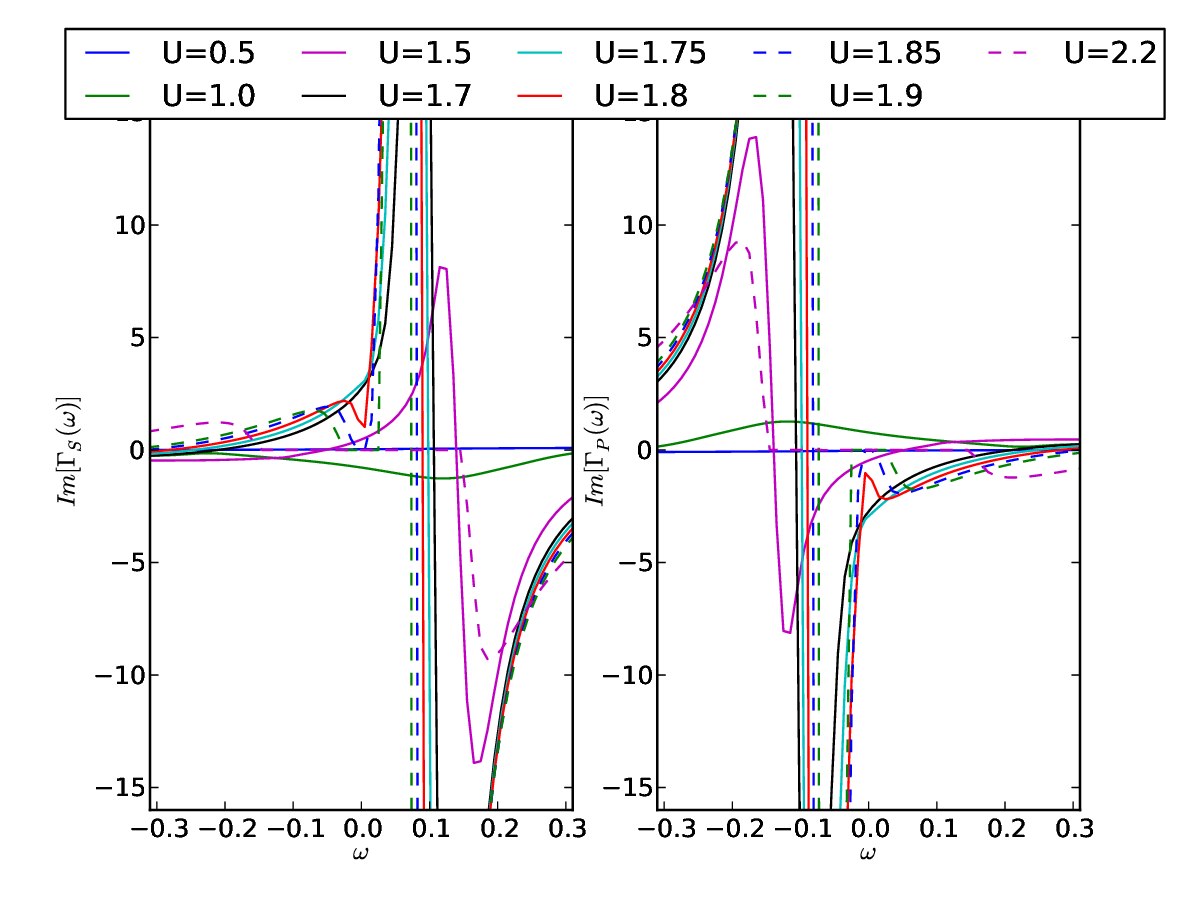} 
\caption{(Color online) Imaginary parts of the irreducible particle-hole vertex functions in the $S,P$ channels as a function of $U$ for the binary-alloy disorder model.  Clear non-analyticities in $\Gamma_{S,P}(\omega)$ at $\omega=0$ occur precisely at $U_{c}=1.8$ (red curve), where the continuous Hubbard band-splitting transition occurs.  In addition, the results confirm the symmetry Im$\Gamma_{S}(\omega)=-$Im$\Gamma_{P}(-\omega)$.}
\label{fig:fig6}
\end{figure}

  In Fig.~\ref{fig:fig6}, we exhibit the imaginary parts of the cluster-momentum-resolved irreducible particle-hole vertex functions as functions of $U$.  
It is clear that both, Im$\Gamma({\bf K},\omega)$ with ${\bf K}=(0,0)$ (called ``S'') and with ${\bf K}=(\pi,\pi)$ (called ``P'') show non-analyticities 
precisely at $\omega=0$ at $U_{c}=1.8$ (red curves).  Thus, for the completely random case, we find, as expected, that the ``Mott'' QCP is signaled by
a clear non-analyticity in the momentum-dependent (irreducible) p-h vertices at the Fermi energy ($\omega=0$).  This non-analytic feature goes hand-in-hand with 
a power-law variation of Im$\Sigma_{S(P)}(\omega)$ in the vicinity of the Fermi energy ($\omega=0$).  Along with spectral functions and self-energies,
the vertex functions also satisfy the ``symmetry'' relation, Im$\Gamma({\bf K}=(0,0),\omega)=-$Im$\Gamma({\bf K}=(\pi,\pi),-\omega)$ for the p-h symmetric case.
Clearly, the anomalous infra-red behavior of the irreducible vertices is directly related to the clear non-analytic structures in the cluster self-energies discussed above.  

   Additional notable features characteristic of $1/d$ effects captured by CDMFT become apparent upon repeating the above procedure for the case of finite
``alloy'' short-range order (SRO), namely, when $f_{0\alpha}=\langle x_{0}x_{\alpha}\rangle -\langle x_{0}\rangle\langle x_{\alpha}\rangle \neq 0$.
\begin{figure}
\includegraphics[width=1.\columnwidth , height= 
1.\columnwidth]{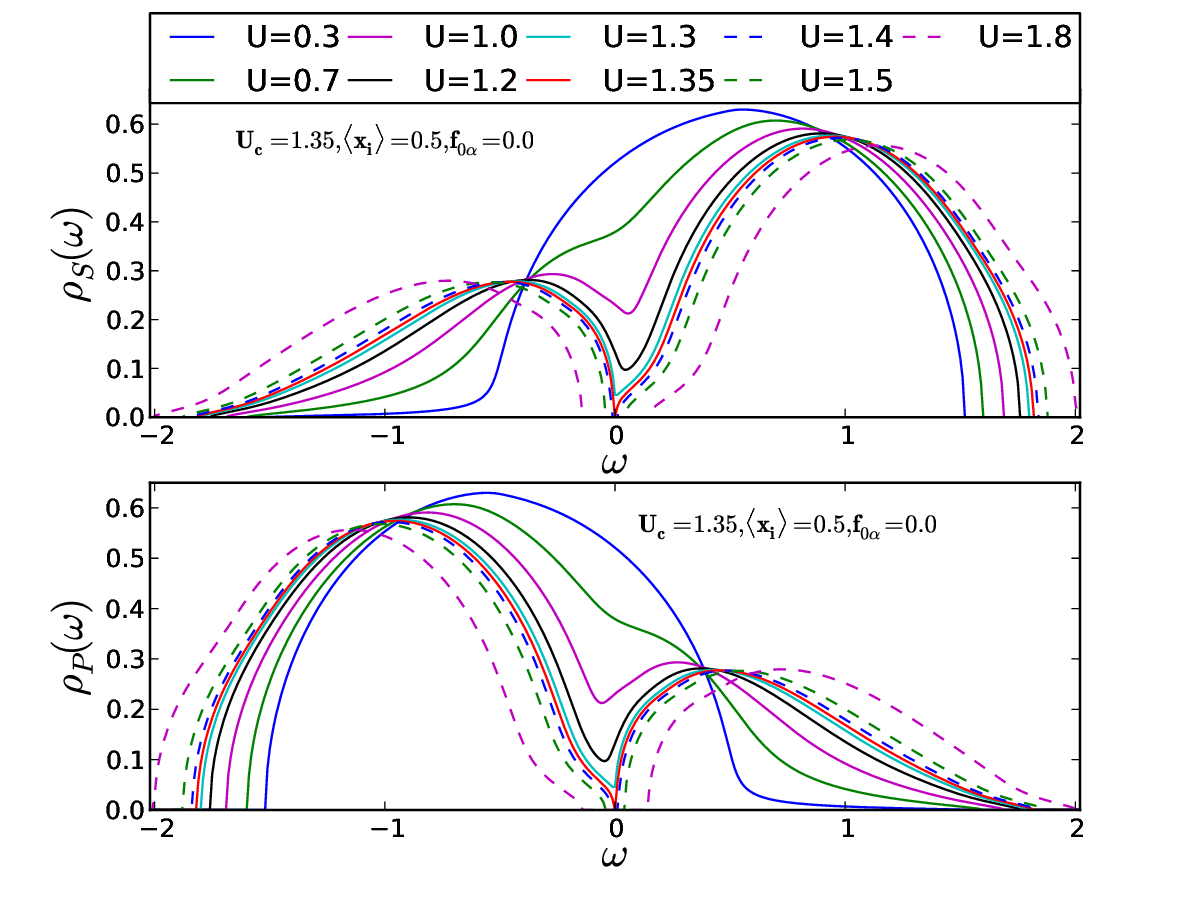} 
\caption{(Color online) Cluster-momentum resolved one-electron spectral functions as a function of $U$ for the short-range ordered binary alloy in the p-h symmetric case.  As expected, the continuous ``Hubbard'' transition now obtains at a smaller $U_{c}=1.35$ (red curve), due to enhanced suppression of itinerance by the ``anti-ferro'' alloy short-range order ($f_{0\alpha} =-0.15<0$).  As for the totally random alloy, the symmetry relation for the spectral functions still holds.  The LDOS shows very similar quantum-critical singular features at low energy at $U_{c}$.}
\label{fig:fig7}
\end{figure}
\begin{figure}
\includegraphics[width=1.\columnwidth , height= 
1.\columnwidth]{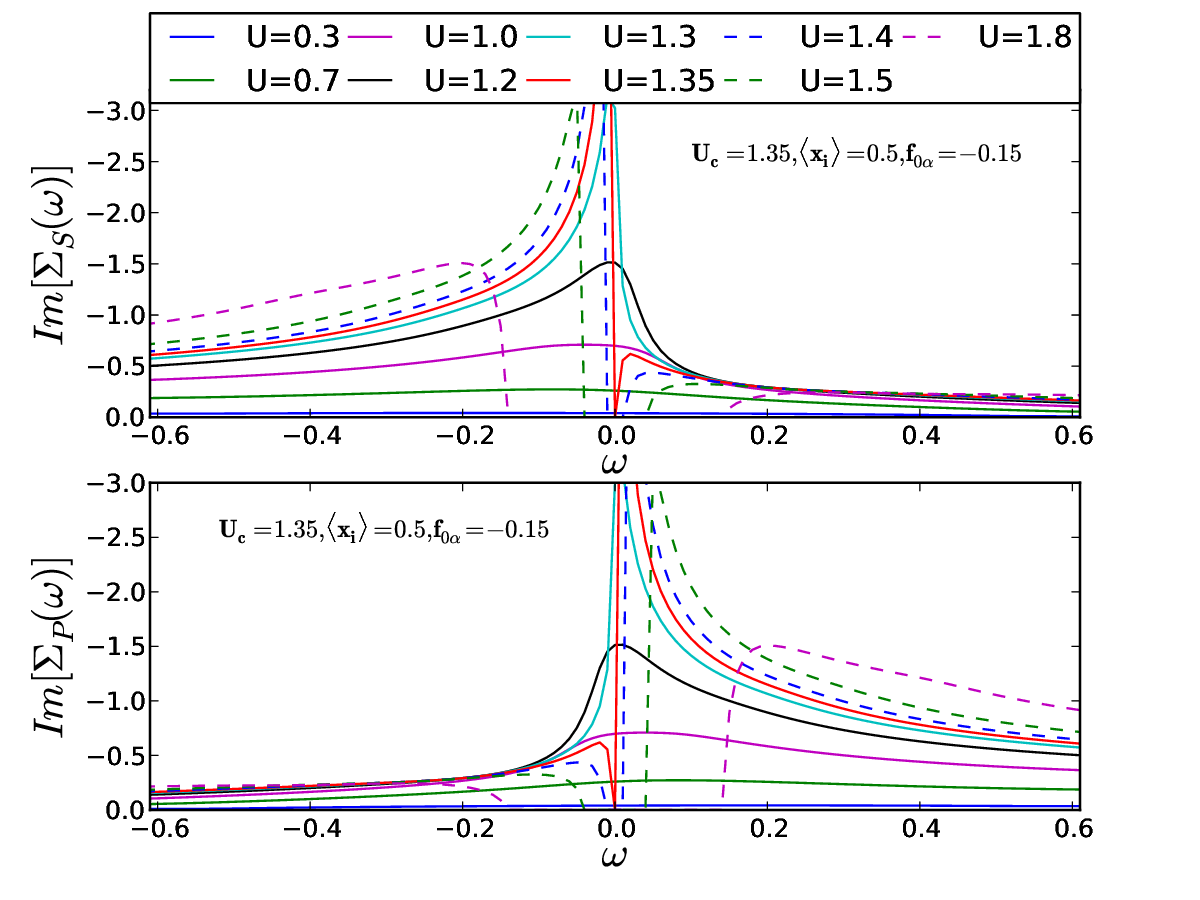} 
\caption{(Color online) Imaginary parts of the cluster-momentum resolved one-particle self-energies as a function of $U$ for the short-range ordered binary alloy.  The symmetry relation for the cluster self-energies still holds, as does the fact that both show critical power-law behavior at $U_{c}$ (see text).}
\label{fig:fig8}
\end{figure}
\begin{figure}
\includegraphics[width=1.\columnwidth , height= 
1.\columnwidth]{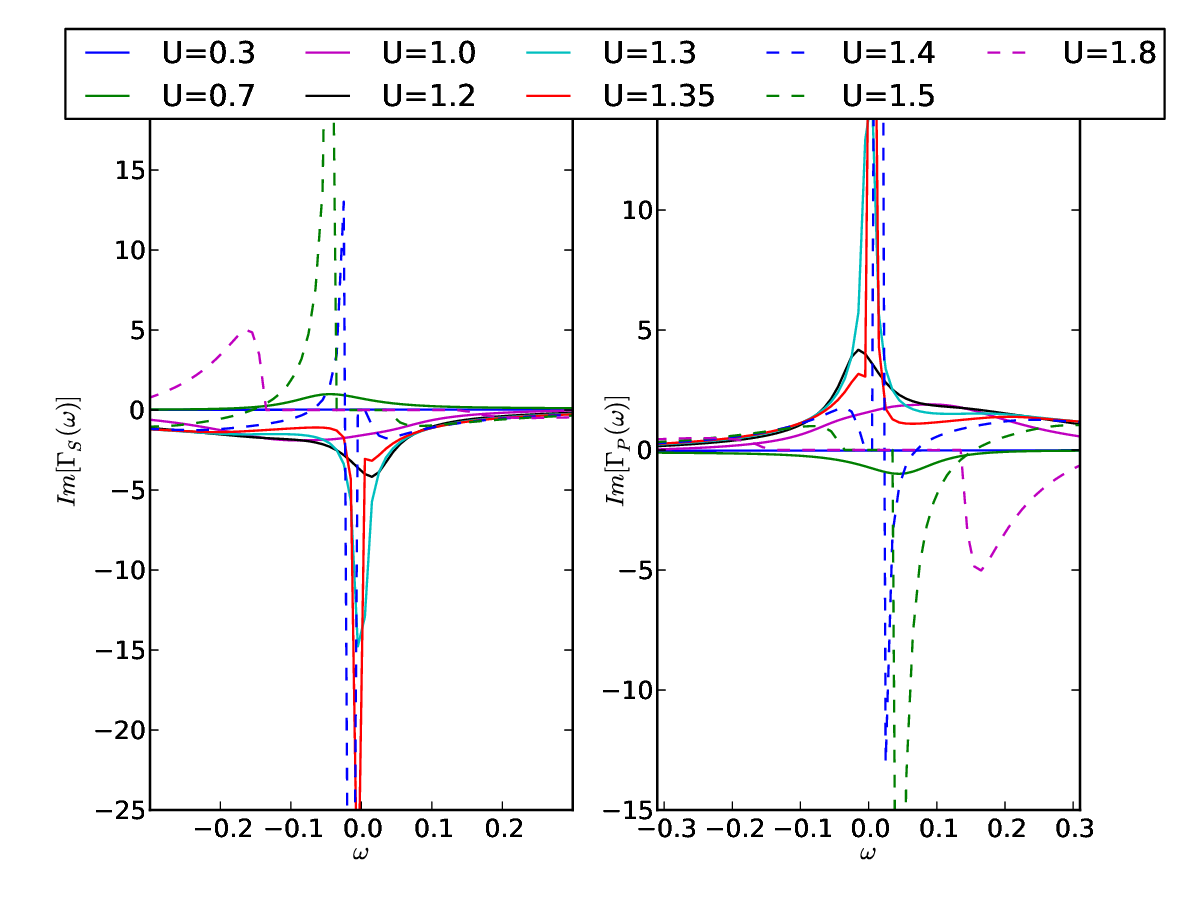} 
\caption{(Color online) Imaginary parts of the cluster-momentum resolved irreducible p-h vertex functions for the p-h symmetric short-range ordered binary alloy as a function of $U$.  Clear non-analyticities in both 
$\Gamma_{S,P}(\omega)$ occur slightly {\it before} the Hubbard-type band-splitting transition occurs, signifying the onset of a novel kind of localization (see text).}
\label{fig:fig9}
\end{figure}

In Fig.~\ref{fig:fig7},  Fig.~\ref{fig:fig8} and Fig.~\ref{fig:fig9}, we exhibit the cluster spectral functions, self-energies and p-h vertices for the case of $f_{0\alpha}=-0.15$, which 
represents the physical situation with short-range ``antiferro'' alloy correlations on the two-site cluster.  Now, the MIT occurs at a critical $U_{c1}=1.35$, 
smaller than for the completely random case.  The reason is simple: on very general grounds, short-ranged ``antiferro'' alloy correlations suppress the 
one-electron hopping by a larger amount compared to the random case (this is also reflected in the deeper pseudogap in the incoherent metal for $f_{0\alpha} <0$),
simply because the probability for an electron to hop onto its neighbor on the cluster is reduced when there is more probability of having a local potential $U$
on the neighboring site.  In this case, Im$\Sigma({\bf K},\omega)$ shows, on first glance, a behavior similar to the case with $f_{0\alpha}=0$ described before.
Upon closer scrutiny of Fig.~\ref{fig:fig8}, however, we find that Im$\Sigma({\bf K},\omega)$ already {\it diverges} for $U=1.3$, slightly {\it before} Hubbard 
band-splitting occurs (cyan curve).  Also, Fig.~\ref{fig:fig8} also clearly shows the power-law divergence of the self-energy (cyan and red curves), with 
Im$\Sigma_{00}(\omega)\simeq |\omega|^{-\eta}$, with $\eta=1/3$ at the MIT.  This new feature is very different from the pole-divergence of the self-energy in the Hubbard model 
within DMFT, but is indeed seen in the DMFT solution for the FKM when the self-energy and the vertex function are treated consistently at the local 
level~\cite{janisjpcm}.  A related non-analyticity in Im$\Gamma({\bf K},\omega)$ also correspondingly occurs at precisely the same value in 
Fig.~\ref{fig:fig9}.  Thus, in this case, we find that the irreducible p-h vertex diverges before the actual MIT occurs.  Such features are also known for the $d=2$ 
Hubbard model within the dynamical vertex approximation~\cite{held}.  However, this divergence of the vertex function is {\it not} associated with 
$(\frac{\partial}{\partial \omega} Re\Sigma_{00}(\omega))_{\omega=0}=\infty$, where the actual ``Mott'' transition occurs.  Thus, it is 
neither connected to any symmetry-breaking (which would require a divergence in the momentum channel), nor does it lead to non-analyticities in the one-particle
response (the LDOS remains smooth for all $U<U_{c1}=1.35$).        
 
   Thus, at the level of spectral functions and self-energies, our CDMFT for the FKM finds universal features at a quantum-critical ``Mott'' transition that are 
qualitatively similar to those found by Janis {\it et al.}~\cite{janisjpcm}  We find that the infra-red non-analytic behavior in $\Gamma({\bf K},\omega)$ {\it precedes} the MIT.
This was probably to be expected, since both approaches deal with quasi-local quantum criticality suited to the Mott-Hubbard problem.  The advantages of our 
extension relate to $(i)$ having a CDMFT that always respects causality~\cite{laad}, and $(ii)$ enables computation of momentum-resolved spectral responses,
even for the hitherto scantily considered cases of explicit ``alloy'' short-range order.  Importantly, having an almost analytic cluster extension of DMFT means 
that we have to simply deal with $N$ coupled non-linear algebraic equations to compute the full CDMFT propagators for a $N$-site cluster, even with short-range
order.  This is an enormous numerical simplification when one envisages its use for {\it real} disordered systems, with or without strong Hubbard correlations:
these issues have long been extremely well-studied using the coherent-potential approximation (CPA) and DMFT~\cite{rmp1996}.  We anticipate wide uses of such a 
semi-analytic approach as ours in this context.

It is interesting to compare our results to those obtained by Shinaoka {\it et al.}~\cite{imada-PRL2009}.  Motivated by disordered and correlated systems near a MIT, they consider a disordered Hubbard model, where Hubbard correlations are treated within static Hartree-Fock, giving rise to local moments, while disorder effects over and above HF are studied by exact diagonalization techniques.  Their main findings are $(i)$ a ``soft'' gap arises even with purely
local interactions, in contrast to that in an Efros-Shklovskii picture, where it arises from long-range coulomb interactions and $(ii)$ while the LDOS
$A(E)\simeq |E-E_{F}|^{\alpha}$ with $0.5<\alpha <1$ for $|E-E_{F}|>0.1$, they see that $A(E)\simeq$ exp$[-(-\gamma$log$|E-E_{F}|)^{3}]$ provides a much better fit for $|E-E_{F}|<0.1$.  In contrast, we find that the LDOS, $\rho(\omega)\simeq C|\omega|^{1/3}$ remains valid up to lowest energies at the QCP: this is similar to the situation found in single-site DMFT~\cite{vandongen}, where precisely the same behavior is found analytically.

   These differences could arise from many factors: $(a)$ there are {\it no} localized magnetic moments in our case, since we do not have the Hubbard term,
$(b)$ while we focus on predominantly short-range disorder correlations, Shinaoka et al include longer range disorder correlations.  It is noteworthy that a ``soft power-law gap'' already appears in (C)DMFT studies, and while it is conceivable that the low-energy behavior may change upon increasing cluster size, this remains to be shown.  Alternatively, if local moments {\it are} crucial to obtain this behavior, one must study the disordered Hubbard model within CDMFT.  This ambitious enterprise is left for future consideration.

\vspace{0.5cm}

{\bf Charge Susceptibility and Response to a Sudden Local Quench}

\vspace{0.5cm}

  In addition to universal critical features found in the last section within an exact-to-$O(1/d)$ CDMFT for the FKM, additional details regarding the nature of
this strong-coupling ``Mott'' transition can be gleaned from examination of the two-particle response.  In particular, the dynamic charge susceptibility of the
FKM can also be precisely computed in our approach by using the CDMFT propagators ($G_{S(P)}({\bf k},\omega)$) and the irreducible p-h vertices $\Gamma_{S(P)}(\omega)$
(notice that the latter have dependence on the cluster momenta ${\bf K}$) in the Bethe-Salpeter equation, as detailed in ``Model and Solution'' section.  
\begin{figure}
\includegraphics[width=1.\columnwidth , height= 
1.\columnwidth]{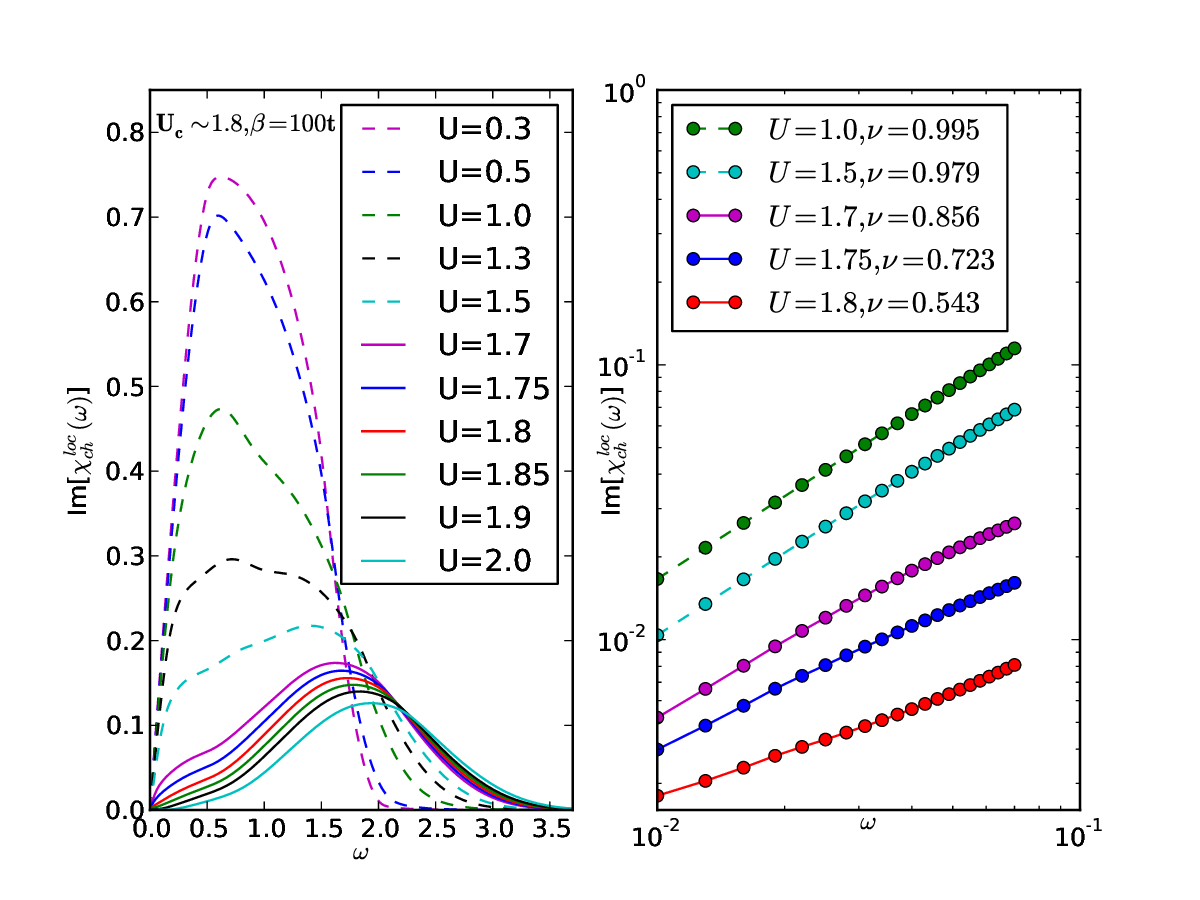} 
\caption{(Color online) The imaginary part of the local component of the full dynamical charge susceptibility for the p-h symmetric binary alloy disorder model in the totally random case ($f_{0\alpha}=0$).  Up to $U_{1}= 1.4$, Im$\chi_{ch}^{loc}(\omega)\simeq \omega$, similar to its DMFT counterpart.  However, for $1.5\leq U\leq U_{c}=1.8$, Im$\chi_{ch}^{loc}(\omega)\simeq \omega^{\nu}$ where $\nu=1-\kappa$ and $0<\kappa(U)<1$ and $\kappa$ 
reduces with increasing $U$, reaching a value $\kappa=0.5$ at $U_{c}$ (red curve).  This has very unusual consequences for the long-time response to a ``sudden'' local quench at strong coupling (see text).}
\label{fig:fig10}
\end{figure}

In Fig.~\ref{fig:fig10}, 
we show the imaginary part of the full cluster-local dynamical charge susceptibility as $U$ increases.  On first glance, our results are quite similar to those in 
earlier DMFT work~\cite{freericks}.  Beginning from small $U$ up to $U\simeq 1.2$, Im$\chi_{ch}(\omega)$ varies linearly with $\omega$ in the infra-red, with a maximum 
at intermediate energy, followed by a high-energy fall-off.  However, closer scrutiny of the strong-coupling ($U\geq 1.4$) regime reveals that this behavior undergoes 
a qualitative change at low energies:  now Im$\chi_{ch}(\omega)\simeq \omega^{\nu}$, with $\nu=1-\kappa$ and $0<\kappa(U) <1$.  It is important to notice that the configurationally
averaged DOS does {\it not} show any non-analyticities in this regime, and the system is close to, but not in the ``Mott'' insulating regime.  A closer look at the 
behavior of the cluster self-energies and irreducible vertex functions in this regime shows that both begin to acquire non-trivial energy dependence at low energy when 
$U$ is close to the critical value needed for the ``Mott'' transition to occur.  In fact, as described before, (see both Im $\Gamma_{S(P)}(\omega)$ start exhibiting strong
$\omega$-dependence, especially near $\omega=0$, when $U\geq 1.4$, and clear non-analyticities accompanied by anomalous power-law variation near $\omega=0$ when one is 
very close to the transition in the range $1.7<U<1.8$.  Thus, it is clear that the anomalous low-energy behavior of the collective charge fluctuations is linked to the 
strong $\omega$-dependence and impending non-analytic behavior in the cluster irreducible vertex as the MIT as approached from the metallic side.  Thus, while the fact that 
the vertex diverges {\it before} the actual MIT does not lead to non-analyticity in the one-electron spectral functions, it does qualitatively modify the collective density fluctuations, 
reflecting in an anomalously overdamped critical form.  We are unaware of such a connection existing within earlier DMFT studies~\cite{freericks} and this qualitatively new feature has 
not previously been noticed, to our best knowledge.

   One interpretation of this unusual feature is the following.  Close to the Hubbard band splitting (``Mott'')
 transition, one generically expects formation of excitons.  A simple way to understand this is in terms of the 
``holon-doublon'' mapping of the model, which is a partial particle-hole transformation where $c^{\dag}\rightarrow c^{\dag}, d\rightarrow d^{\dag}$.  Now $U\sum_{i}n_{i,c}n_{i,d}\rightarrow -U\sum_{i}(n_{i,c}n_{i,d}- n_{i,c})$, whereby the $c,d$ fermions experience an on-site {\it attraction}, leading to formation of local ``pair'' bound states (these are excitons in the original model) of the type $\langle c_{i}^{\dag}d_{i}^{\dag}\rangle$.  Quite generally, in a Hubbard model, one expects these bosons to Bose-condense.  An 
upshot thereof is the well-known fact that this is nothing else but antiferromagnetic magnetic order, now 
interpreted as a Bose condensate of spin excitons.  In our simplified FKM or binary-alloy case, however, such
a BEC is explicitly forbidden by the fact that the {\it local} $Z_{2}$ gauge symmetry, associated with 
$[n_{i,d},H]=0$ for all $i$, cannot be spontaneously be broken by Elitzur's theorem.  This still leaves open
the possibility of having inter-site excitonic pairing of the $c$-fermions on the two-site cluster.  Without global broken symmetry, such a state would be a dynamically fluctuating excitonic liquid. One would
expect that a phase transition to a ``solid'' of such excitonic pairs will eventually occur, perhaps as a 
Berezinskii-Kosterlitz-Thouless (BKT) transition~\cite{kopec}, but this is out of scope of the present work.
However, having strong inter-site excitonic liquid fluctuations could cause the irreducible charge vertices to exhibit precursor features, and it could be that our finding above is a signal of such an impending instability.  More work is certainly needed to put this idea on a stronger footing, but this requires a separate investigation.  

   Finally, one would expect emergence of anomalous features in vertex functions and charge fluctuations close to the MIT to have deeper ramifications.  Specifically, we now 
address the question outlined in the Introduction:  ``Can we study the long-time response of the FKM to a sudden local quench, and can such an endeavor provides deeper insight into the
``strong'' localization aspect inherent in a continuous ``Mott'' transition?''  In other words, if we introduce a local, suddenly switched-on potential in the manner of a deep-core hole
potential in metals, how would the long-time response of the ``core-hole'' spectrum evolve with $U$?  In the famed instance of a Landau Fermi liquid metal, the seminal work of 
Anderson~\cite{pwa1967}, Nozieres and de Dominicis~\cite{1969} (AND) leads to the result that at long times, the core-hole propagator, related to the wave function overlap between the
ground states without and with the suddenly switched potential, goes like a power-law: $\rho_{h}(t)|_{t\rightarrow\infty}\simeq t^{-\alpha}$ with $\pi\alpha=$tan$^{-1}(V_{h}\rho_{c}(0))$
being the ($s$-wave for a local scalar potential) scattering phase shift.  It has also long been shown that~\cite{mh1971} the deep reason for this feature is that the particle-hole 
fluctuation spectrum, $\rho_{ph}(\omega)$ (related to the collective charge fluctuation response), in a Fermi gas is linear in energy.  Explicit evaluation of the core-hole response
when $e-e$ interactions in the Landau Fermi liquid sense are present is a much more involved and delicate matter~\cite{janisjpcm}.  It is clear that qualitative change(s) in the low-energy
density fluctuation spectrum must qualitatively modify the long-time response to such a sudden quench.  

   Answering this question in our case of the FKM is a subtle matter, since the $c$-fermion spectral function is {\it not} that of a Landau Fermi liquid, but describes an incoherent 
non-Landau Fermi liquid state.  As long as Im$\chi_{ch}(\omega)\simeq \omega$ holds, however, we expect that the long-time response will be similar to that evaluated by Janis~\cite{janisjpcm}
using rather formal Wiener-Hopf techniques.  Ultimately, the long-time response still behaves in a qualitatively similar way to that for the free Fermi gas, except that the exponent in the
power law is modified by interactions (thus, Im$\chi_{ch}(\omega)\simeq \omega$ still holds, but with sizable renormalization).  In our case, we thus expect that $\rho_{h}(t)|_{t\rightarrow\infty}\simeq
t^{-\alpha}$ still holds for $U<1.3$, since we do find Im$\chi_{ch}(\omega)\simeq \omega$ in this regime in the infra-red.  However, the qualitative change to the form 
Im$\chi_{ch}(\omega)\simeq \omega^{1-\kappa}$ with $0<\kappa <1$ in the infra-red for $U\geq 1.4$ must also qualitatively modify the long-time overlap and the ``core-hole'' response.

   Rather than resort to a direct computation of the long-time response within CDMFT, we will find it more instructive to consider this issue by using the low-energy results gleaned from CDMFT 
as inputs into an elegant approach first used in the context of the seminal X-ray edge problem by Schotte and Schotte ~\cite{schotte} and by M\"uller-Hartmann {\it et al}~\cite{mh1971}.  To this end, we have to 
identify the collective charge fluctuations encoded in $\chi_{ch}(\omega)$ with a bath of bosonic particle-hole excitations in the incoherent metal.  Generally, using the linked cluster expansion,
 the spectral function of the localized ``core-hole'' is

\be
S_{h}(\omega)=\frac{1}{2\pi}\int_{-\infty}^{\infty} dt e^{i\omega t} exp[V_{h}^{2}\int_{0}^{\infty} dE Im \chi_{ph}(E)\frac{e^{-iEt}-1}{E^{2}}]
\ee
where $V_{h}$ is the ``suddenly switched'' core-hole potential.  As long as Im$\chi_{ph}(E) \simeq E$, we estimate, similar to the well-known result, that the core-hole spectral function behaves like
$ S_{h}(\omega)\simeq \omega^{-\alpha}$ with $\alpha=(1/\pi)tan^{-1}(V_{h}\rho_{00}(0))$, with $\rho_{00}(0)$ being the CDMFT LDOS at the Fermi energy (in a full computation, this exponent will change a
bit because $\rho_{00}(\omega)$ has sizable frequency dependence close to $\omega=0$ at strong coupling in the metal as found in Results, but the qualitative features will survive).  However, when 
$U \geq 1.4$, having Im $\chi_{ch}(\omega)\simeq \omega^{1-\kappa}$ must modify this well-known behavior.  In this regime we find (see also Ref.~\cite{mh1971}) the following leading contribution to the
core-hole spectral function

\be
S_{h}(\omega) \simeq \frac{V_{h}^{2}}{E_{F}}(\frac{E_{F}}{\omega})^{1+\kappa} exp[-\pi V_{h}^{4}(\frac{E_{F}}{\omega})^{2(1-\kappa)}]
\ee
which is qualitatively distinct from the well-known form, and corresponds to a long-time wave function overlap having a very non-standard form: $\rho_{h}(t)|_{t\rightarrow\infty}\simeq e^{-t^{1-\kappa}}$.
This qualitative modification of the long-time wave function overlap is a strong manifestation of a novel type of localization at work.  It would be tempting to associate this with a many-body localized regime, especially since the Landau quasiparticle picture is also violated within this strong-coupling regime, but more work is called for to clinch this issue.  The basic
underlying reason for this novel behavior is the same as the one leading to generation of the anomalous exponent $\kappa$ in the p-h fluctuation spectrum, $i.e$, strong $\omega$-dependence 
and incipient non-analyticity in the irreducible p-h vertex close to the MIT.
         
\vspace{1.0cm}

{\bf DISCUSSION AND CONCLUSION}

\vspace{0.5cm}

Using the disordered binary alloy analogy extended to a two-site cluster, we have investigated $1/d$ effects on the continuous MIT in the "simplified" FKM (by this, we mean a FKM where the disorder is quenched, rather than annealed, so quantities like $\langle x_{0}\rangle$ and $f_{0\alpha}$ are fixed and given from a binary
distribution, rather than computed self-consistently, as in the true FKM).  In spite of this simplification,
we find that quantum critical features at the level of one-electron Green functions and self-energies are very similar to those obtained from an ``Anderson-Falicov-Kimball''~\cite{janisjpcm} model.  This is not so surprising, since the effect of the FK term, $U\sum_{i}n_{id}n_{ic}$ is precisely to generate a band ``splitting'' for all
$U$ in the FKM as well, and a binary alloy disorder indeed has exactly a similar effect on the DOS.  Thus, 
within DMFT or cDMFT approaches such as ours, one would expect quantitative changes in the spectral functions, but no qualitative modification of critical exponents in the LDOS exactly at the band-splitting Hubbard-like transition.  

  However, in strong contrast to the one-electron response, inclusion of {\it non-local} irreducible p-h vertex in computation of the dynamic charge susceptibility does lead to qualitatively new effects at strong coupling.  We have shown that Im$\chi_{ch}(\omega)\simeq \omega^{1-\kappa}$ with $0<\kappa <1$ to occur precisely in the same regime where the non-local vertex shows strong frequency dependence and signs of an impending non-analyticity (the latter occurs either at the MIT, or precedes it, see above).  This feature is quite anomalous, indicating that a novel collectively fluctuating state of the electronic fluid, characterized by infra-red critical {\it bosonic} p-h modes, sets in before the MIT occurs.  Naturally, one expects that this feature will drastically modify the charge responses in the strong coupling limit: in fact, related effects should reveal themselves in optical response of the disordered electron fluid.  We leave detailed elucidation of such points for future work.

  To summarize, we have analyzed the role of short-ranged (spatially non-local) alloy correlations on the Hubbard-like MIT in a binary disorder Anderson model at strong coupling in detail.  While quantum critical 
features at one-electron level are exactly similar to recent DMFT results~\cite{janisjpcm} for the disordered FKM, non-local vertex corrections show up rather dramatically as a qualitative change in character of collective
p-h spectrum at strong coupling.  In contrast to previous CPA studies, this is a concrete manifestation of the relevance of dynamical effects associated with $1/d$ alloy correlations near the quantum critical point 
associated with a continuous MIT of the Hubbard band-splitting type.  It is obviously of interest to elucidate the nature and consequences of this strong coupling QCP in various transport responses.  This aspect is under study, and will be reported separately.   
   
\vspace{0.5cm}

{\bf Acknowledgements}  We are grateful to Prof. T. V. Ramakrishnan for insightful discussions.

\end{document}